\documentclass[%
reprint,
superscriptaddress,
 amsmath,amssymb,
 pre,
]{revtex4-1}

\usepackage[normalem]{ulem}
\usepackage{graphicx}
\usepackage{array}
\usepackage{array}
\usepackage{tabularx}
\usepackage{dcolumn}
\usepackage{bm}
\usepackage{booktabs}

\usepackage[utf8]{inputenc}
\usepackage[T1]{fontenc}
\usepackage{mathptmx}
\usepackage{etoolbox}
\usepackage{verbatim}
\usepackage{float}
\usepackage{caption}
\usepackage{subcaption}
\usepackage{amsmath}
\usepackage{amssymb}
\usepackage{xcolor}
\usepackage{hyperref}
\usepackage{cleveref}
\usepackage{booktabs}
\usepackage{booktabs}
\usepackage{tabularx}
\usepackage{multirow}
\usepackage{array}
\usepackage{hyperref}

\begin{document}


\title{Quasi mono-energetic, relativistic electron acceleration in a femtosecond, high intensity laser excited solid magnet.}

\author{Trishul Dhalia}%
\email{trishuldhalia@gmail.com}
\affiliation{Indian Institute of Technology Delhi, Hauz Khas, New Delhi 110016, India}

\author{Anandam Choudhary}
\affiliation{Tata Institute of Fundamental Research, 1 Homi Bhabha Road, Colaba, Mumbai 400 005, India}

\author{C. Aparajit}
\affiliation{Tata Institute of Fundamental Research, 1 Homi Bhabha Road, Colaba, Mumbai 400 005, India}

\author{Amit D. Lad}
\affiliation{Tata Institute of Fundamental Research, 1 Homi Bhabha Road, Colaba, Mumbai 400 005, India}

\author{Ankit Dulat}
\affiliation{Tata Institute of Fundamental Research, 1 Homi Bhabha Road, Colaba, Mumbai 400 005, India}

\author{Yash M. Ved}
\affiliation{Tata Institute of Fundamental Research, 1 Homi Bhabha Road, Colaba, Mumbai 400 005, India}

\author{Rohit Juneja}
\affiliation{Indian Institute of Technology Delhi, Hauz Khas, New Delhi 110016, India}

\author{Amita Das}
\email{amita@iitd.ac.in}
\affiliation{Indian Institute of Technology Delhi, Hauz Khas, New Delhi 110016, India}

\author{G Ravindra Kumar}
\email{grk@tifr.res.in}
\affiliation{Tata Institute of Fundamental Research, 1 Homi Bhabha Road, Colaba, Mumbai 400 005, India}

\date{\today}
\begin{abstract}

The interaction of ultraintense lasers with magnetized overdense plasmas reveals a fundamentally new regime of laser-driven particle acceleration. Particle-in-cell simulations demonstrate the generation of directional, quasi-monoenergetic electrons of 2 MeV with $\sim 6.7\%$ energy spread, superimposed on a broad thermal electron background with the estimated acceleration gradient of $3.6 ~\textrm{MeV}/\mu m$, which is the highest till date. In contrast to conventional laser-plasma accelerators, which rely on underdense plasmas and are therefore constrained to relatively low plasma densities and limited beam charge, the present scheme operates in plasmas with densities millions $(\times10^6)$ of times higher, opening new possibilities for the generation of high-flux energetic electron beams. A central result of this work is the demonstration of the excitation of electron Bernstein waves during relativistic laser interaction with magnetized overdense plasmas. The subsequent Landau damping of these electrostatic warm-plasma modes selectively transfers energy to resonant electrons, leading to the emergence of quasi-monoenergetic spectral peaks at energies that can be tuned through the applied magnetic field. To support the simulation results, we experimentally demonstrate the directional emission of energetic electrons from a simple permanent-magnet target irradiated by an ultraintense laser pulse, highlighting the practical feasibility of controlled electron-beam generation in dense plasma environments. These findings establish electron Bernstein waves as an efficient mediator of laser energy coupling in overdense plasmas and introduce a new paradigm for controlled particle acceleration and energy deposition in high-energy-density plasma systems.
\end{abstract}

\maketitle
\section{Introduction}

Large fluxes of relativistic electrons are typically generated with solid targets. Several innovative methods have increased the fluxes,, namely the use of nanostructured surfaces \cite{rajeev_PhysRevLett.90.115002,Purvis2013}, sub-lambda gratings \cite{kahaly_PhysRevLett.101.145001, Lad2022, Macchi_10.1063/1.5013321, Fedeli_PhysRevLett.116.015001} etc. The high background density that the generated electron beam experiences, however, poses several complexities. The beam divergence,  shown to increase with laser intensity \cite{Green_PhysRevLett.100.015003} is an important parameter that needs to be controlled for the effective transfer of energy to secondary emissions. The beam is also prone to several instabilities in the dense surrounding plasma \cite{drake_2006,Weibel_PhysRevLett.2.83,CalifanoPhysRevLett.96.105008,CalifanoPhysRevLett.86.5293} and this can seriously damage its integrity via filamentation in as short a transport length as a few tens of microns, shown in several computer simulations. In the initial stages of transport, the particles may be treated as ballistic but with loss of energy and increasing divergence in the transport, collisional as well as collisionless \cite{yabuuchi2009evidence,Tikhonchuk10.1063/1.1459061,ManclossiPhysRevLett.96.125002,ShengPhysRevLett.85.5340} stopping can play a significant role. In applications like fast ignition of laser fusion, \cite{Tabak_10.1063/1.870664, Gong2019} petawatt peak power, femtosecond/picosecond laser-driven mega-ampere electron beams are required to deposit their energy in a very small (tens of micrometers) region, propagating through an imploded super-solid density target \cite{Vauzour_PhysRevLett.109.255002}.   This is an issue that continues to challenge us.

How can we guide these mega fluxes of electrons through a solid?  One approach has been to use tailored targets.  For example, carbon nanotubes \cite{Chatterjee_PhysRevLett.108.235005} and metal nanochannels \cite{PKSingh_PhysRevSTAB.16.063401} have been employed to enhance the generation as well as transport of large currents over lengths much larger than the typical filamentation lengths of a few microns. An interesting innovation has been the use of the resistivity gradient in the medium surrounding the plasma channel, called resistive collimation \cite{Bell_PhysRevLett.91.035003}, to guide the electrons \cite{Kar_PhysRevLett.102.055001, Ramakrishna_PhysRevLett.105.135001}, going so far to maximize energy deposition in a specified ignition volume \cite{Robinson_PhysRevLett.108.125004}. Tailoring the density gradient has also been suggested as a possible way to guide electrons \cite{yadav2009anomalous}. While all these are internal to the target, large external magnetic fields (10$^3$ tesla) have been postulated in simulations \cite{Cai_10.1063/1.4812631, Bailly-Grandvaux2018,WangPhysRevLett.114.015001} and even experimentally applied to imploded fusion cores \cite{Sakata2018} to reduce divergence and enhance coupling of the laser to the electrons. The availability of increasingly high magnetic fields in the laboratory in fact, leads to an interesting and altogether new magnetized regime of laser plasma interactions for which theoretical and numerical studies are being carried out  \cite{das2020laser, PhysRevE.105.055209,vashistha2020new,goswami2021ponderomotive}.

 The production and application of highly monoenergetic MeV and GeV electron beams via laser wakefield acceleration (LWFA) and plasma wakefield acceleration (PWFA)  is a great achievement of relativistic laser-plasma interaction research \cite{modena1995electron, mangles2004monoenergetic,geddes2004high, faure2004laser,pukhov2002laser,zhang2025plasma}. The earliest demonstrations of laser--plasma accelerators produced quasi-monoenergetic electrons in the MeV energy range \cite{andreev2000monoenergetic, tsung2006simulation,krushelnick2005laser}, while subsequent advances in ultra-intense laser technology and plasma guiding techniques significantly reduced the energy spread, improved collimation, and pushed energies to the GeV range \cite{PhysRevLett.111.165002,wang2013quasi,shaw2025path}. However, a fundamental limitation of these schemes arises from the requirement that the laser propagate through the plasma, which is possible only in the underdense regime. This imposes an inherent constraint on the plasma density and consequently limits the achievable accelerating gradients as well as the total charge and current.

There is therefore a strong motivation to explore a fundamentally different paradigm capable of overcoming some or all of these limitations. In conventional overdense plasmas, the laser interaction is restricted primarily to the plasma boundary because electromagnetic waves cannot propagate into the dense medium \cite{chen1984introduction}. Recent studies have shown that applying strong external magnetic fields can open propagation pass bands in the magnetized plasma dispersion relation, thereby enabling laser penetration into overdense plasmas. However, such studies have so far shown only energy transfer from the laser to plasma particles, resulting in a broad thermal distribution of electrons and ions \cite{dhalia2023harmonic,dhalia2025laser,dhalia2026pushing,juneja2023ion,vashistha2020new,kumar2019excitation,mandal2021electromagnetic,PhysRevE.105.055209,dhalia2024absorption}.

 In the present work, we demonstrate that it is in fact possible to generate quasi-monoenergetic electrons in overdense plasmas using externally applied magnetic fields. More importantly, the mechanism does not require extremely strong magnetic fields capable of modifying electromagnetic wave dispersion, nor does it require laser propagation deep into the overdense medium. Remarkably, the plasma itself generates the required strong magnetic fields from a tiny external magnetic field  \cite{choudhary202510}. Further, the laser-plasma interaction at the surface itself excites electron Bernstein waves \cite{bernstein1958waves,laqua2003electron}, whose subsequent Landau damping gives rise to distinct resonant peaks in the electron energy spectrum. The identification and prediction of electron Bernstein wave excitation in relativistic laser-plasma interactions is itself highly significant, as these electrostatic waves can propagate in overdense magnetized plasmas where ordinary electromagnetic modes cannot penetrate. Their excitation, therefore, opens a fundamentally new avenue for channeling laser energy into dense plasma media and for enabling the controlled generation of energetic particles in regimes previously considered inaccessible to conventional laser-plasma acceleration schemes.  To support the above assertions, we report experiments indicating that even relatively modest externally applied magnetic fields are sufficient, since a dynamo-like mechanism operating during the interaction amplifies the magnetic field to strengths necessary for the process \cite{choudhary202510}.

The manuscript is organized as follows. In Section II, we describe the simulation model and the parameters employed in the study. Section III presents the observations demonstrating the generation of monoenergetic electrons and identifies their origin in the excitation and subsequent damping of electron Bernstein waves. In Section IV, we demonstrate the experimental feasibility of the proposed mechanism using a table-top laser interacting with a simple magnetized target, resulting in the generation of energetic electron beams.

\section{Simulation Details}\label{sec:simulation}
\begin{figure}
    \centering
    \includegraphics[width=3.5in]{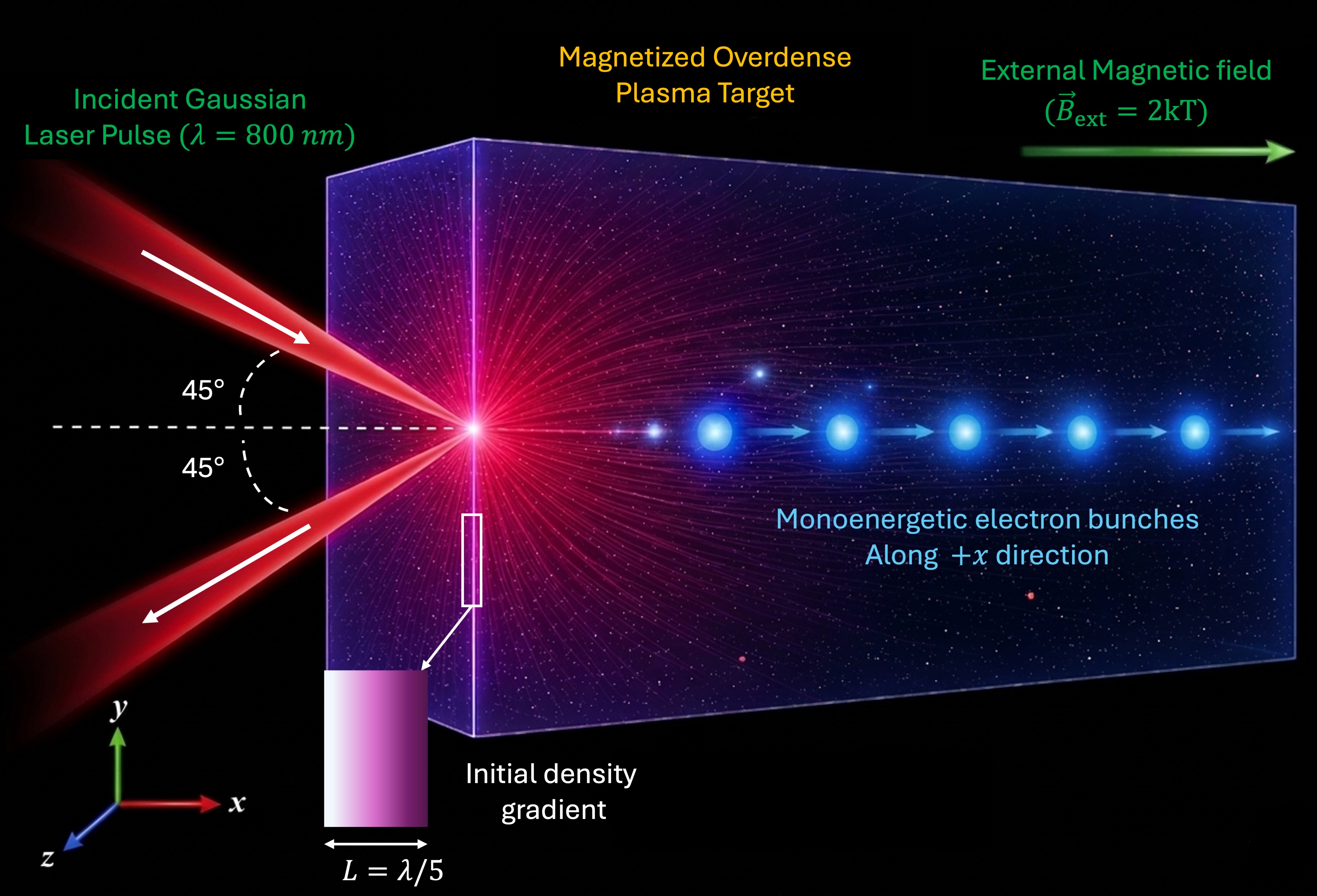}
    \caption{Schematic (not to scale) shows that the simulation geometry chosen in our simulation for a high-intensity laser interacting with a highly overdense plasma surface at an oblique $45^{\circ}$ angle in the presence of an external magnetic field. The direction of the external magnetic field is normal to the plasma surface .}
    \label{fig:schematic_5}
\end{figure}

We performed two-dimensional, three-velocity (2D3V) Particle-In-Cell (PIC) simulations using the OSIRIS 4.0 framework \cite{fonseca2002osiris,hemker2000particle}. A p-polarized laser pulse with wavelength $\lambda = 800~\mathrm{nm}$ (frequency $\nu_L = 3.7 \times 10^{14}Hz$), was chosen to be incident obliquely from the left boundary onto the plasma surface at an angle of $45^\circ$. The plasma target is assumed to be fully ionized and overdense, with an electron density of $100 n_c$. Thus, the angular laser frequency is related to the plasma frequency by  $\omega_l = 0.1\omega_{pe}$.  The target plasma consists of electrons in a stationary neutralizing ion background, with the ions assumed to remain immobile throughout the simulation. The simulation domain size is chosen to be  ($1000 c/\omega_{pe}  \times 1000 c/\omega_{pe}$). The laser pulse has a duration of $5$ laser cycles, corresponding to a temporal width of $\Delta \tau = 500\omega_{pe}^{-1}$ ($21.5~\mathrm{fs}$), while the transverse spot size of the laser is  $250 c/\omega_{pe}$ ($3.6~\mu\mathrm{m}$). For most of the simulations, the normalized vector potential of $a_0 = 6.1$, corresponding to a laser intensity of $8 \times 10^{19}~\mathrm{W/cm^2}$,  has been employed.  An exponentially sharp density gradient was introduced at the front surface of the target having the profile given by the following expression:   \[n=100n_c(\exp((x-y)\ln(2)/L)-1)\] Where the density scale length  $L = 12 c/\omega_{pe}$ ($152~\mathrm{nm} =\lambda/5$). Absorbing boundary conditions are applied at all simulation boundaries. A schematic representation and additional details of the simulation setup are provided in Fig.~\ref{fig:schematic_5}. In addition, a magnetic field $\vec{B}_{ext}$ normal to the target surface is applied.

\begin{figure*}
    \centering
    \includegraphics[width=1\linewidth]{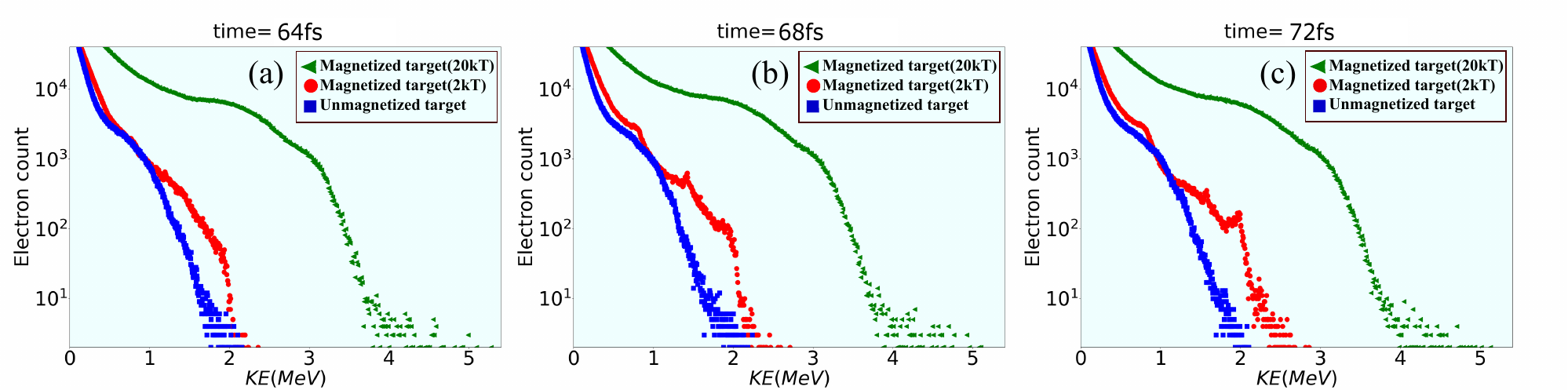}
    \caption{Figure (a,b,c) demonstrates the energy distribution of electrons in the whole simulation box for three different targets (non-magnetic, magnetic $(2kT)$, and magnetic $(20kT)$ at a later stage, when the laser has left the simulation box.}
    \label{fig:dist_with_0_2_20}
\end{figure*}
\begin{figure}
    \centering
    \includegraphics[width=1\linewidth]{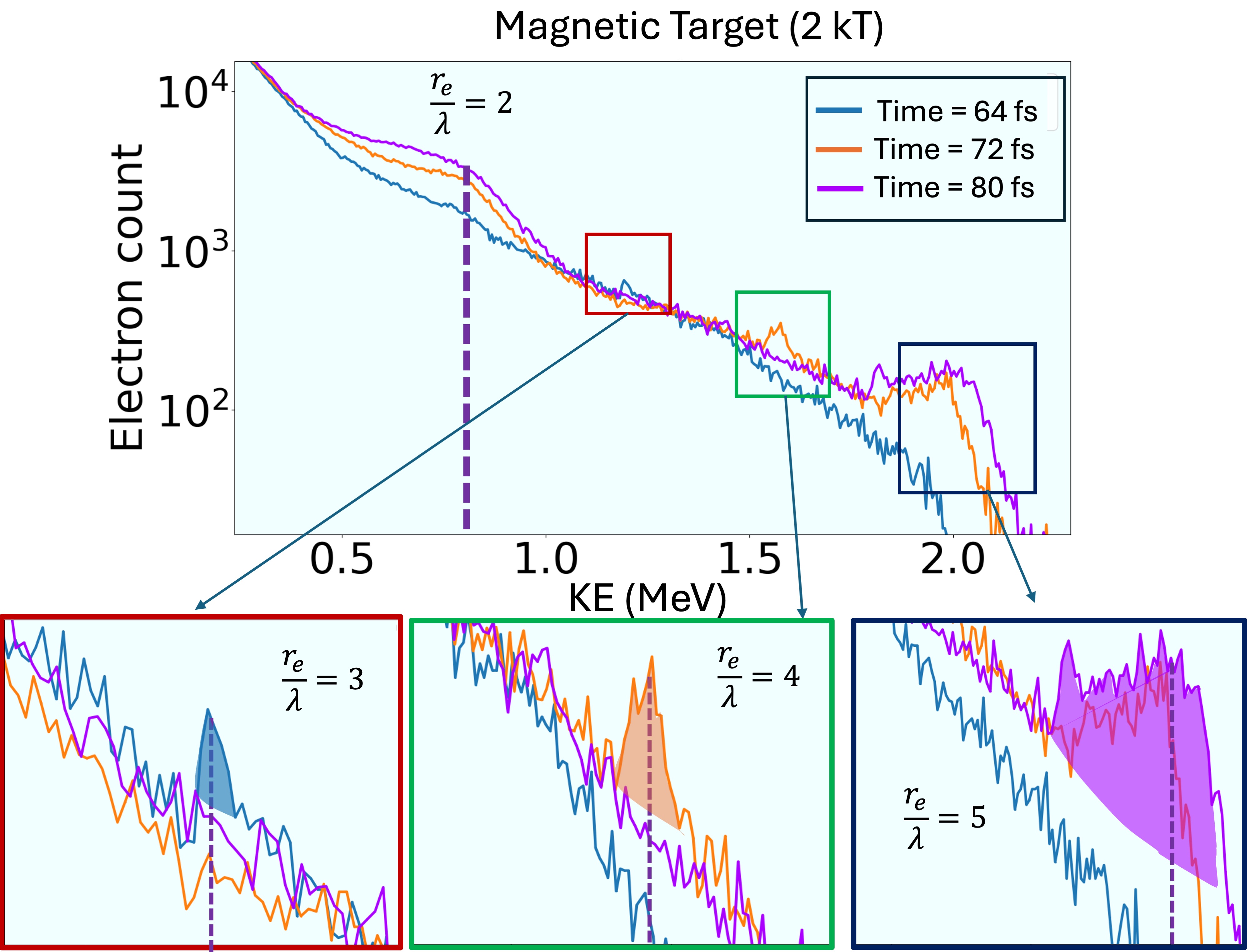}
    \caption{Peaks in the energy spectrum at the locations of integer values of relativistic modified $r_e/\lambda$ ratio. The shaded region shows the population of non-thermal monoenergetic electrons in the distribution.}
    \label{fig:mono_energetic}
\end{figure}
\begin{figure}
    \centering
    \includegraphics[width=1\linewidth]{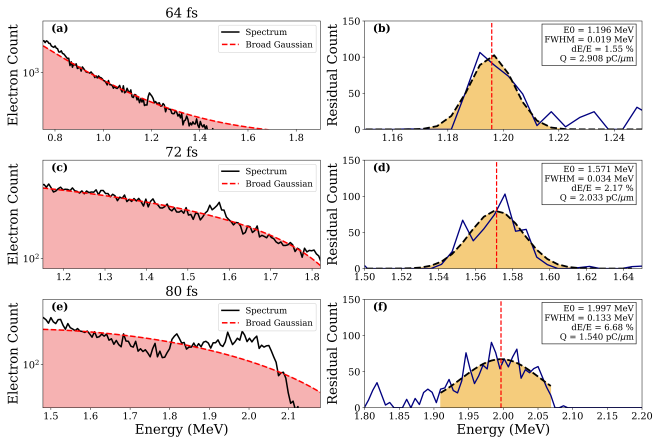}
    \caption{Figure $(a,c,e)$ presents energy distribution of 2kT magnetized target, fitted locally with Gaussian distributions at various times, and figure $(b,d,f)$ shows the corresponding residual (Spectrum-Broad Gaussian) non-thermal population of electrons.}
    \label{fig:Full_monoenergetic}
\end{figure}

\section{Observations and Analysis}\label{sec:observation}
Simulations were primarily performed for  several distinct values of the externally applied magnetic field.  A choice of $B_{ext}=0$, $2~\mathrm{kT}$, and $20~\mathrm{kT}$, referred to hereafter as cases (i), (ii), and (iii), respectively.  The  cyclotron frequency $\omega_{ce}$,  for these cases are (i) $0$ (ii) $3.52 \times 10^{14}~\mathrm{rad/s}$ and (iii) $3.52 \times 10^{15}~\mathrm{rad/s}$ respectively. Thus, case (i) corresponds to the interaction of the laser with an unmagnetized overdense plasma target and serves as the reference configuration. In case (ii), the applied magnetic field is sufficiently weak that the laser propagation remains essentially governed by the unmagnetized plasma dispersion relation; however, the post-absorption dynamics of the electrons are strongly influenced by the magnetic field. In contrast, case (iii) represents a strongly magnetized regime in which the applied field is large enough to significantly modify the electromagnetic dispersion characteristics of the plasma. As a result, the laser can penetrate the overdense target by propagating through the pass band of the magnetized plasma dispersion relation.

Figure~\ref{fig:dist_with_0_2_20} presents a comparison of the electron energy spectra for the three cases at different stages of the interaction. It is evident that the maximum electron energy increases with the strength of the externally applied magnetic field, indicating enhanced laser-to-electron energy coupling in the magnetized configurations. For most cases, the electron spectra remain broad and exhibit the characteristic features of a thermal distribution. However, a striking departure from this behavior is observed in case (ii) ($B_{ext}=2~\mathrm{kT}$), where distinct peaks emerge in the energy spectrum at later times. Remarkably, these spectral peaks develop well after the laser pulse has completely exited the simulation domain, indicating that the underlying energization process is not a direct consequence of the instantaneous laser field. Instead, it points to a secondary plasma-mediated mechanism that continues to transfer energy to the electrons long after the laser–plasma interaction has ceased. The appearance of these well-defined spectral features thus provides strong evidence for a non-thermal acceleration process that preferentially energizes electrons at specific resonant energies.

We now examine the underlying resonant mechanism responsible for the emergence of these spectral peaks. In particular, we seek to understand the energies at which the peaks appear and the physical process that selects these preferred values. A careful analysis reveals that the peak locations are not arbitrary; rather, they satisfy a well-defined resonance condition involving the laser wavelength, $\lambda$, and the relativistically modified electron gyroradius, $r_e$. Specifically, the spectral peaks occur at energies for which $r_e/\lambda = n$, where $n$ is an integer, as is evident from Fig.~\ref{fig:mono_energetic}. This observation strongly suggests a resonant interaction between the electron gyromotion and the laser-driven plasma dynamics, leading to preferential energy deposition at discrete electron energies.

It should be emphasized that the quasi-monoenergetic peaks emerge on top of a broad thermal electron background. In order to quantify the charge and current content associated specifically with the monoenergetic component, it is necessary to separate the resonant peak from the underlying thermal distribution. The procedure adopted for this decomposition is illustrated in Fig.~\ref{fig:Full_monoenergetic}, where the thermal background is identified and subtracted from the total electron spectrum. The remaining excess electron population can then be attributed to the resonantly accelerated electrons responsible for the spectral peak. Using the electron count contained within this residual peak, we estimate the corresponding beam charge and energy spread are 1.540 $pC/\mu m$ and 6.68 $\%$ respectively for the highest QME electron beam of $\sim 2$ MeV. These values provide a quantitative measure of the effectiveness of the proposed acceleration mechanism in generating directed quasi-monoenergetic electron bunches.

The above wavelength-based resonance condition can be readily recast in terms of frequencies. Using the relation between the electron gyroradius and the relativistically modified cyclotron frequency $\omega_{ce,R}$, one finds that the resonance condition is equivalently expressed as
\begin{equation}
\frac{\nu_L}{\omega_{ce,R}} = \frac{\nu_L \gamma}{\omega_{ce}} =  n 
\label{eq:Resonance}
\end{equation}
with $\gamma$ being the relativistic factor.
Here $\nu_L$ is the laser frequency and $\omega_{ce}$ is the relativistically corrected electron gyrofrequency. We have approximated the electron speed by $c$. Thus, the observed spectral peaks correspond to harmonic cyclotron resonances, which occur when the laser frequency is an integer multiple of the relativistic electron gyrofrequency.

With this understanding, a coherent picture of the monoenergetic electron generation mechanism emerges. The first requirement is the formation of a sufficiently energetic thermal electron population. This occurs naturally during the laser–plasma interaction, where the component of the laser electric field normal to the plasma surface extracts electrons from the target. At relativistic intensities, the laser-induced $\vec{J}\times\vec{b}_L$ ($\vec{b}_L$ being the laser field) force also contributes to electron extraction and energization.. In the presence of an externally applied magnetic field, the additional force $\vec{J}\times\vec{B}_{ext}$ acts primarily along the target surface, modifying the subsequent electron dynamics and transport.

The combined action of these mechanisms, together with the sheath fields that develop as electrons are pulled out of the plasma, results in the generation of a broad thermal distribution of energetic electrons. In this sense, the initial heating stage may be viewed as a magnetized extension of the well-known vacuum-heating and $\vec{J}\times\vec{B}$-heating processes. The externally applied magnetic field plays a particularly important role in this phase, as evidenced by the systematic increase in electron energies with increasing magnetic-field strength. This enhanced energization provides a larger population of electrons capable of participating in the subsequent resonant interaction responsible for the formation of the quasi-monoenergetic spectral peaks.

 The presence of this hot electron population enables the excitation of warm plasma modes, among which the electron Bernstein wave (EBW) plays a central role here. The excitation of the Bernstein wave clearly depends on  the presence of an external magnetic field. Consequently, a fraction of the laser energy is transferred into the EBW during the interaction. As the thermal electron distribution evolves, electrons whose energies satisfy the resonance condition interact strongly with the EBW. The subsequent Landau damping of the EBW preferentially deposits energy into these resonant electrons, leading to the formation of distinct peaks in the energy spectrum and the emergence of quasi-monoenergetic electron bunches from an initially broad thermal background.

Having outlined the proposed mechanism, we now turn to the simulation results to examine its validity. We seek direct numerical evidence supporting the sequence of processes leading from laser energy deposition to the formation of quasi-monoenergetic electron populations. 

As is evident from Fig.~\ref{fig:dist_with_0_2_20}(a), the electron energy distribution at $t=64~\mathrm{fs}$ is broad for all three values of the applied magnetic field, with no discernible spectral peaks appearing in any of the cases. The maximum electron energy increases with increasing magnetic field strength, indicating enhanced energization in the magnetized configurations. This trend is consistent with the physical picture outlined earlier, wherein the magnetic field promotes more efficient electron heating and energy coupling during the laser–plasma interaction. Nevertheless, the overall spectra at this stage remain predominantly thermal in nature. A qualitatively different behavior emerges at later times. As shown in Fig.~\ref{fig:dist_with_0_2_20}(b) and (c), corresponding to $t=68~\mathrm{fs}$ and $t=74~\mathrm{fs}$, respectively, distinct and well-defined peaks develop in the electron energy spectrum exclusively for case (ii) ($B_{ext}=2~\mathrm{kT}$), while the spectra for cases (i) and (iii) continue to remain broad and essentially featureless.

\begin{figure}[ht!]
    \centering
    \includegraphics[width=1\linewidth]{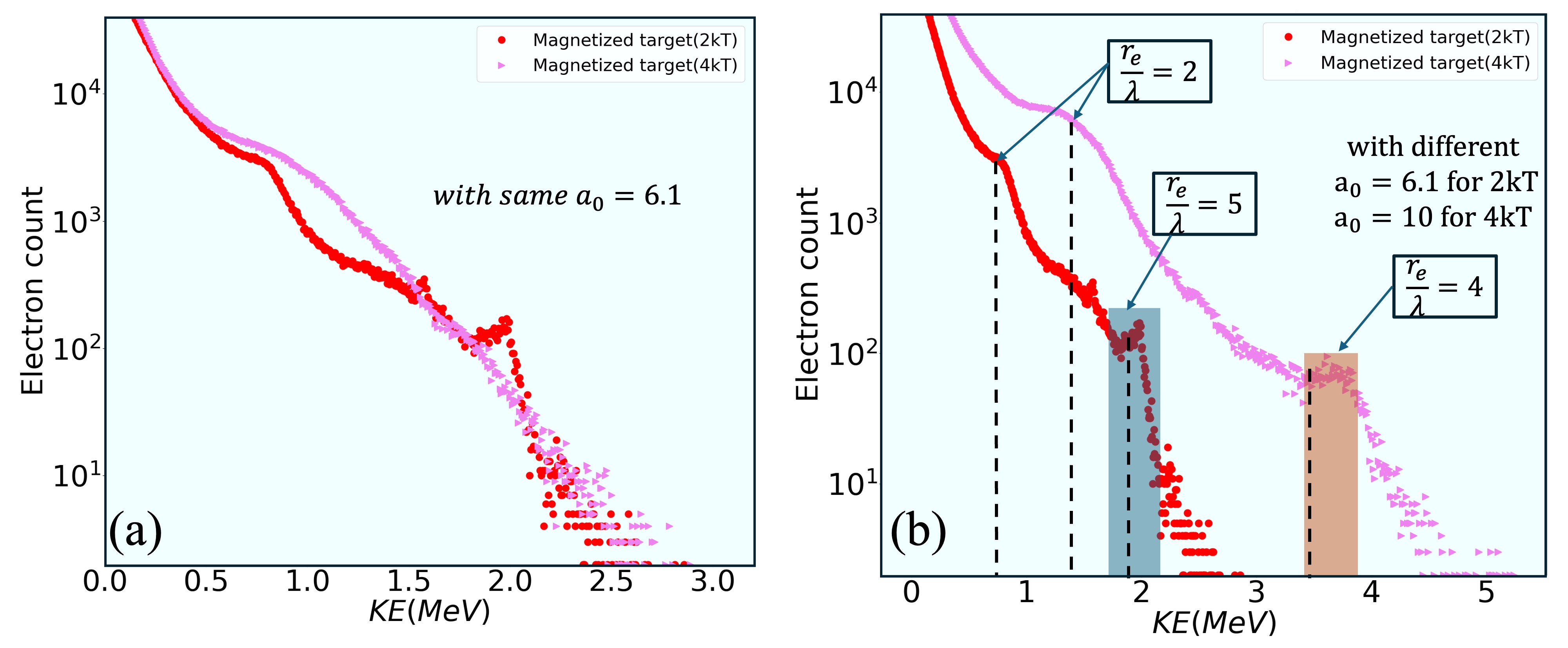}
    \caption{Distribution of electron at 72 fs with (a) same intensity on both magnetic targets, (b) different intensity on both magnetic targets.}
    \label{fig:distribution_2_4_increased_a0}
\end{figure}

\begin{figure*}
    \centering
    \includegraphics[width=5in]{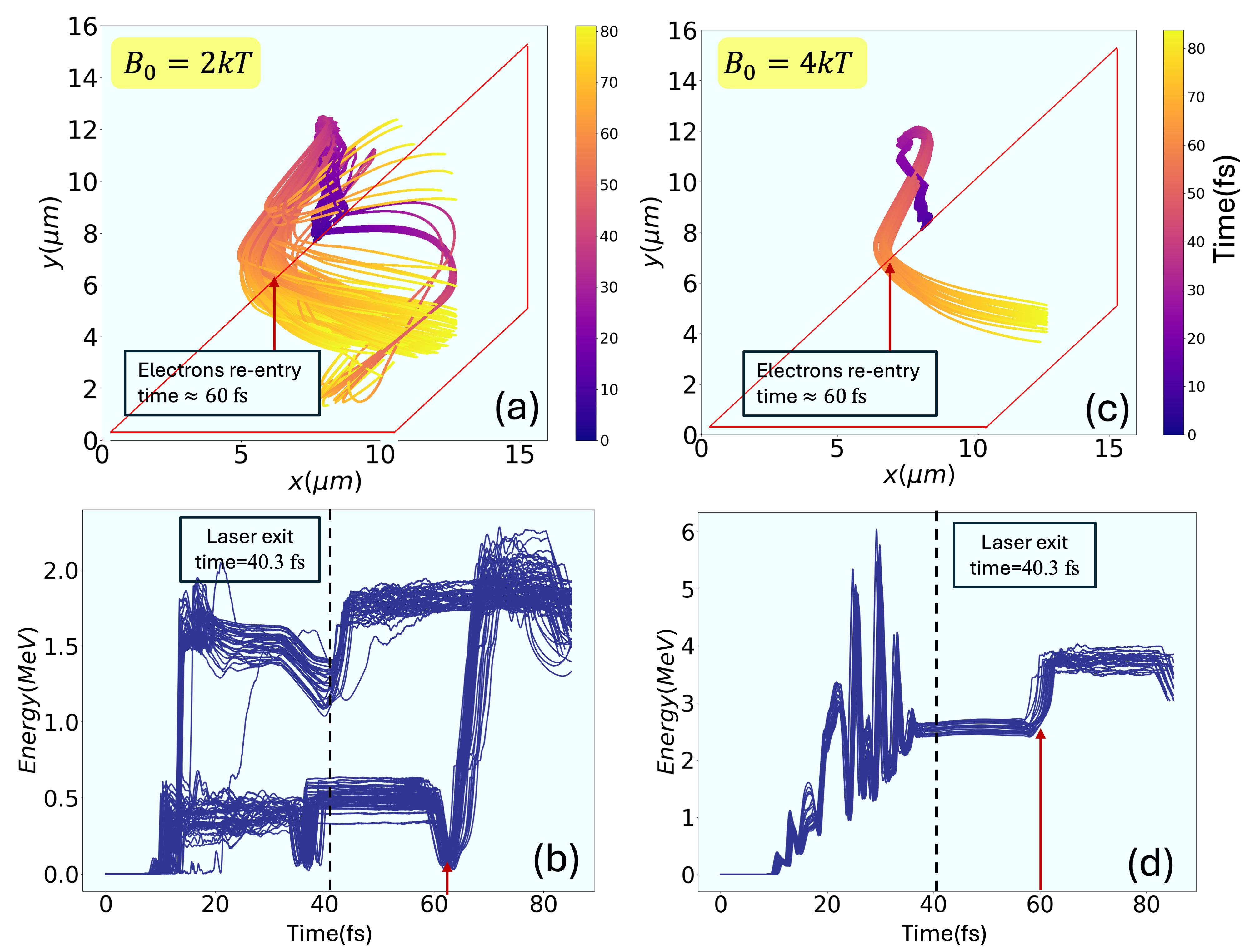}
    \caption{Figure (a,c) shows trajectories of quasi-monoenergetic electrons (from the highlighted region in Fig.~\ref{fig:distribution_2_4_increased_a0}(b)) for magnetic targets of 2 and 4 kT, respectively. Figure (b,d) shows the temporal evolution of energy for tracked electrons in both cases. }
    \label{fig:particle_trajectory}
\end{figure*}

\begin{figure}
    \centering
    \includegraphics[width=1\linewidth]{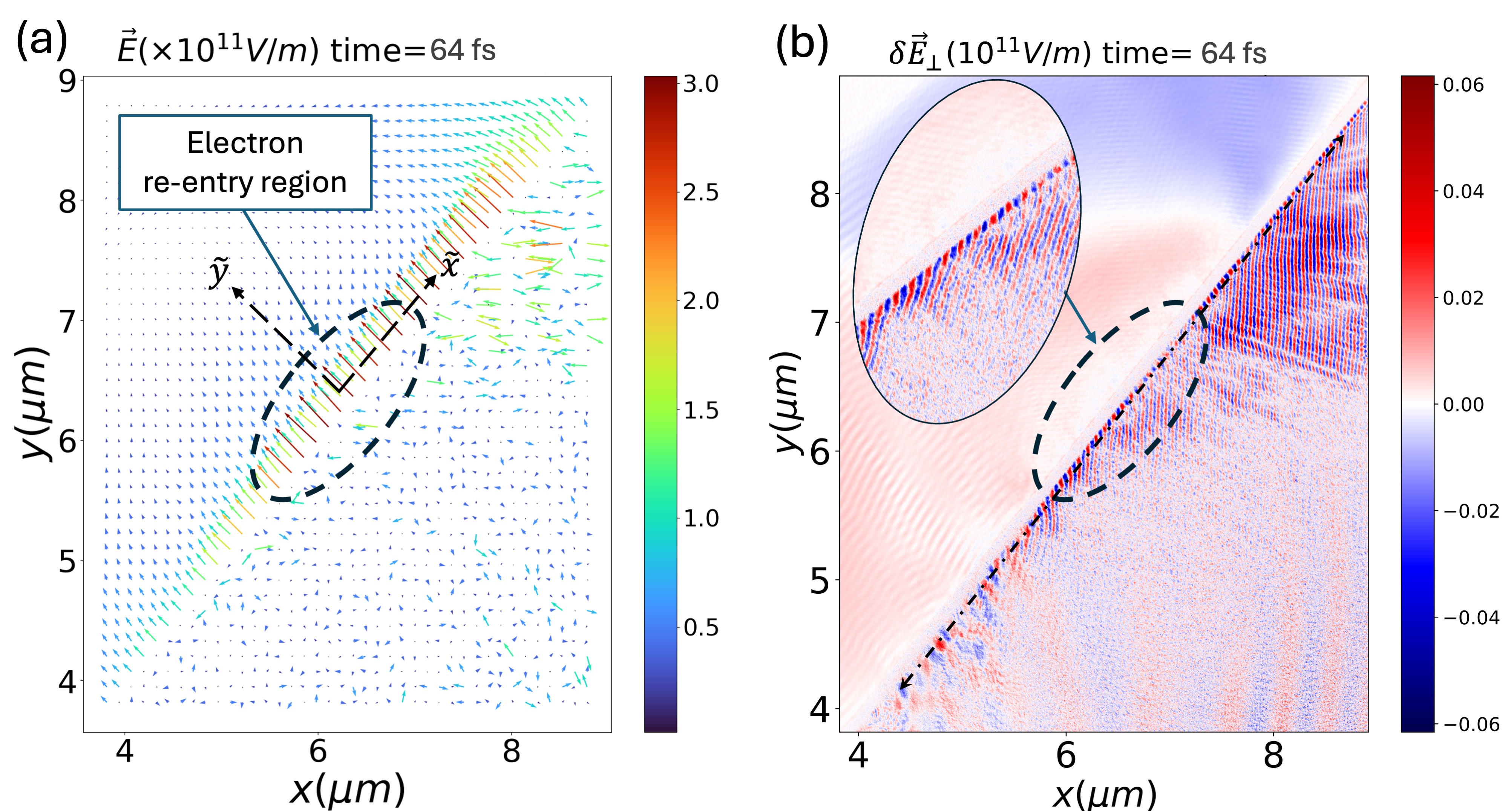}
    \caption{Figure (a) shows a quiver plot of the in-plane electric field $(\vec{E}=E_x\hat{i}+E_y\hat{j})$  and (b) depicts the oscillatory electric field component along the plasma surface subtracted with the average static field component $(\delta\vec{E}_\perp)$ for magnetic target (2kT). }
    \label{fig:sheath_fluctuation}
\end{figure}
\begin{figure}
    \centering
    \includegraphics[width=1\linewidth]{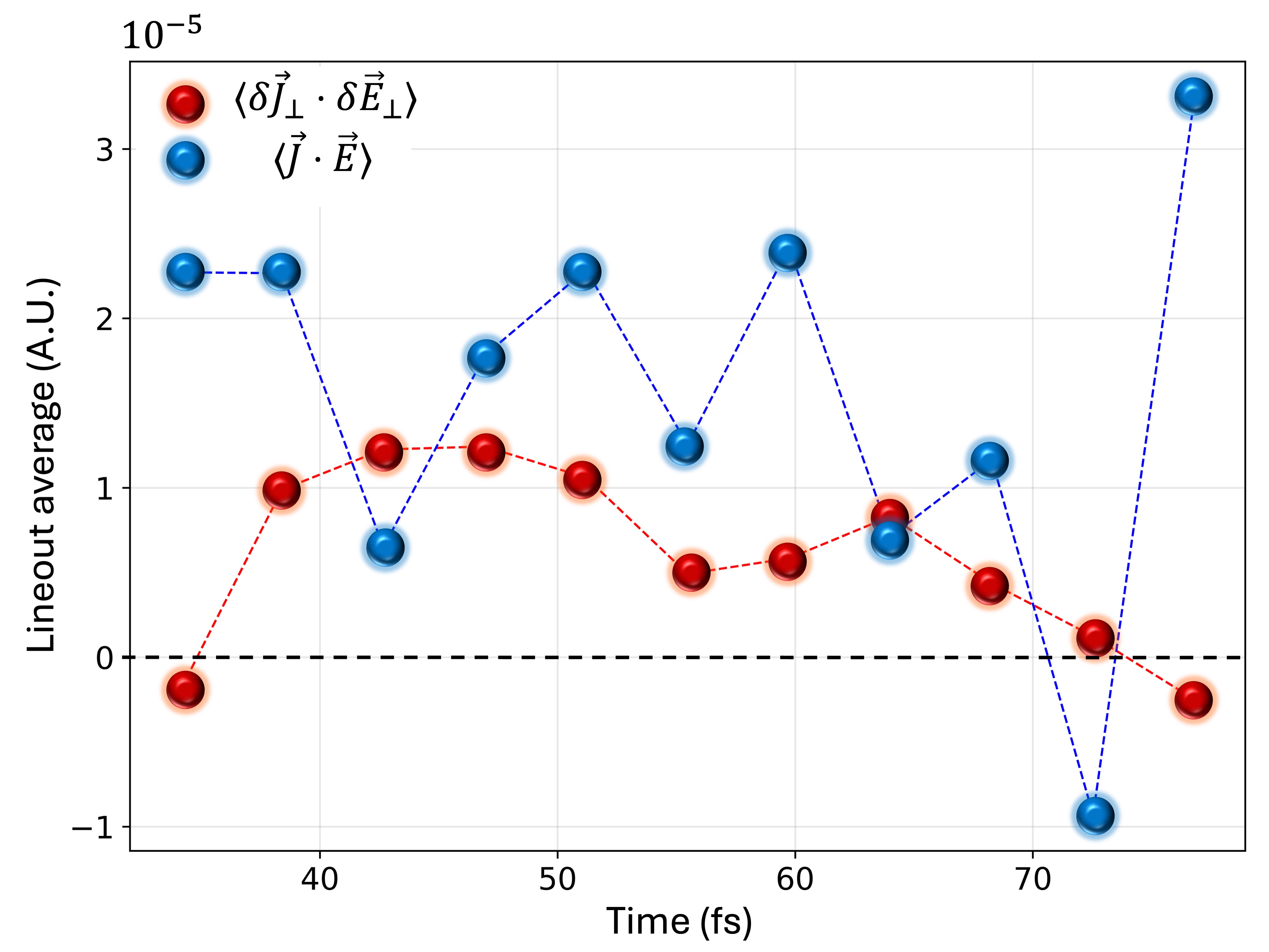}
    \caption{Figure plots the time evolution of lineout average  (indicated by black dotted line in figure \ref{fig:sheath_fluctuation}(b)) of the quantity $\langle\delta \vec{J}_\perp\cdot \delta\vec{E}_\perp\rangle$ and $\langle\vec{J}\cdot\vec{E}\rangle$ with time for magnetic target (2kT).}
    \label{fig:JE_vs_time}
\end{figure}
This behavior can also be readily understood within the framework of the proposed physical model. For the unmagnetized configuration, case (i), the resonance condition described by Eq.~\ref{eq:Resonance} cannot be satisfied since the electron cyclotron frequency is absent. Consequently, no resonant energy transfer mechanism exists to produce distinct spectral peaks. In case (iii), although the plasma is strongly magnetized, the laser frequency $\omega_{L}$ is nearly an order of magnitude smaller than the electron cyclotron frequency $\omega_{ce}$. As a result, even the fundamental resonance ($n=1$) would require the cyclotron frequency to be reduced by a relativistic factor of approximately $\gamma \simeq 10$. Such highly relativistic electrons are not present in significant numbers within the observed energy spectrum, even for the strongest applied magnetic field considered here. Therefore, the resonance condition remains inaccessible in practice, preventing the formation of monoenergetic electron peaks despite the enhanced overall electron heating.

To further validate the proposed mechanism, we performed an additional set of simulations with an external magnetic field of $B_{ext}=4~\mathrm{kT}$. Figure~\ref{fig:distribution_2_4_increased_a0}(a) compares the electron energy spectra obtained for $B_{ext}=2~\mathrm{kT}$ and $4~\mathrm{kT}$ at $t=72~\mathrm{fs}$. In contrast to the $2~\mathrm{kT}$ case, no monoenergetic peak is observed for $B_{ext}=4~\mathrm{kT}$. This behavior is readily explained by the resonance model. Since the cyclotron frequency is twice as large for the $4~\mathrm{kT}$ field, satisfying the resonance condition requires approximately twice the relativistic factor, and hence substantially higher electron energies. The thermal electron population generated under the original laser conditions is insufficient to reach the required resonance energy. To test this prediction, we carried out a further simulation for $B_{ext}=4~\mathrm{kT}$ with an increased laser intensity corresponding to $a_0=10$. As expected, the higher laser intensity produces a more energetic electron population, extending the spectrum to significantly larger energies. Most importantly, under these conditions, a distinct monoenergetic peak reappears at the energy predicted by the resonance condition. The recovery of the spectral peak upon increasing the electron energy provides strong evidence in support of the proposed resonance mechanism and places the physical interpretation on a firmer footing.

In Fig.~\ref{fig:particle_trajectory}, we track the trajectories of electrons belonging to the vicinity of the resonant spectral peak. A comparison of the trajectories shown in subplots (a) and (b), corresponding to $B_{ext}=2~\mathrm{kT}$ and $4~\mathrm{kT}$, respectively, reveals a marked difference in the degree of phase-space localization of these particles. In the presence of the stronger magnetic field, the electron trajectories exhibit significantly tighter bunching, indicating a more coherent resonant interaction. This enhanced confinement is consistent with the stronger magnetic control exerted on the electron motion. Furthermore, subplots (c) and (d) of the same figure reveal that the dominant energy gain of these electrons occurs each time they traverse the plasma boundary. Rather than being accelerated continuously throughout their trajectories, the particles experience discrete bursts of energization during successive crossings of the plasma–vacuum interface. This observation highlights the crucial role of the boundary region in the resonant acceleration process and suggests that the interaction responsible for the energy gain is localized near the plasma surface.  

To gain further insight into the resonant acceleration process, we now examine the electromagnetic field structure in the vicinity of the plasma–vacuum boundary. Such an analysis can help identify the plasma mode responsible for the resonant energy transfer, as its presence should manifest through characteristic spatial field patterns. It is important to note, however, that any such mode develops in the presence of a strong background sheath field that dominates the boundary region. The existence of this sheath field is evident from Fig.~\ref{fig:sheath_fluctuation}(a), which displays the electric-field quiver plot. As expected, the electric field is predominantly oriented normal to the plasma surface, a characteristic signature of the sheath. The electron Bernstein wave (EBW), on the other hand, possesses an oscillatory electric field component perpendicular to the applied magnetic field. Since the external magnetic field in our configuration is directed normal to the plasma–vacuum interface, the EBW electric field is expected to lie primarily in the plane of the boundary. We therefore search for signatures of such a field structure. A direct examination of the  electric field,
$\vec{E}_{\perp} = E_x\hat{i} + E_y \hat{j}$ is not sufficient because a finite laser spot size can produce large-scale transverse variations in the sheath field, leading to a nonzero average component that may mask the comparatively weak oscillatory fluctuations associated with the EBW. To isolate these fluctuations, we subtract from  $\vec{E}_{\perp}$ the line average along the plasma boundary $\langle \vec{E}_{\perp} \rangle$ and analyze only the residual field. The resulting field distribution is shown in Fig.~\ref{fig:sheath_fluctuation}(b). A clear short-wavelength oscillatory structure, oriented perpendicular to the applied magnetic field, is observed. Such a pattern is consistent with the expected characteristics of an electron Bernstein wave and provides strong evidence for its excitation at the plasma surface.

Figure~\ref{fig:JE_vs_time} shows the temporal evolution of the energy transfer rates from field to the particles associated with the interaction. We plot both the total energy transfer rate, quantified by $\langle \vec{J}\cdot\vec{E}\rangle$, and the contribution arising from the damping of the electron Bernstein wave, measured through the fluctuating component $\langle \delta\vec{J}_{\perp}\cdot\delta\vec{E}_{\perp}\rangle$. The latter quantity isolates the energy exchange associated with the wave fields and therefore serves as a direct indicator of EBW-mediated energization. A particularly noteworthy observation is that during the time interval in which the monoenergetic peaks emerge in the electron spectrum, $\langle \delta\vec{J}_{\perp}\cdot\delta\vec{E}_{\perp}\rangle$ remains positive. This indicates a net transfer of energy from the EBW to the electrons.  The temporal correlation between the positive EBW damping rate and the appearance of the spectral peaks provides strong evidence that the resonant electron acceleration is driven by the damping of the electron Bernstein wave. The resonant nature suggests that the transfer mechanism is routed through a Landau damping process, precisely as predicted by the proposed mechanism.

\begin{figure}[!ht]
   \centering
    \includegraphics[width=0.9\linewidth]{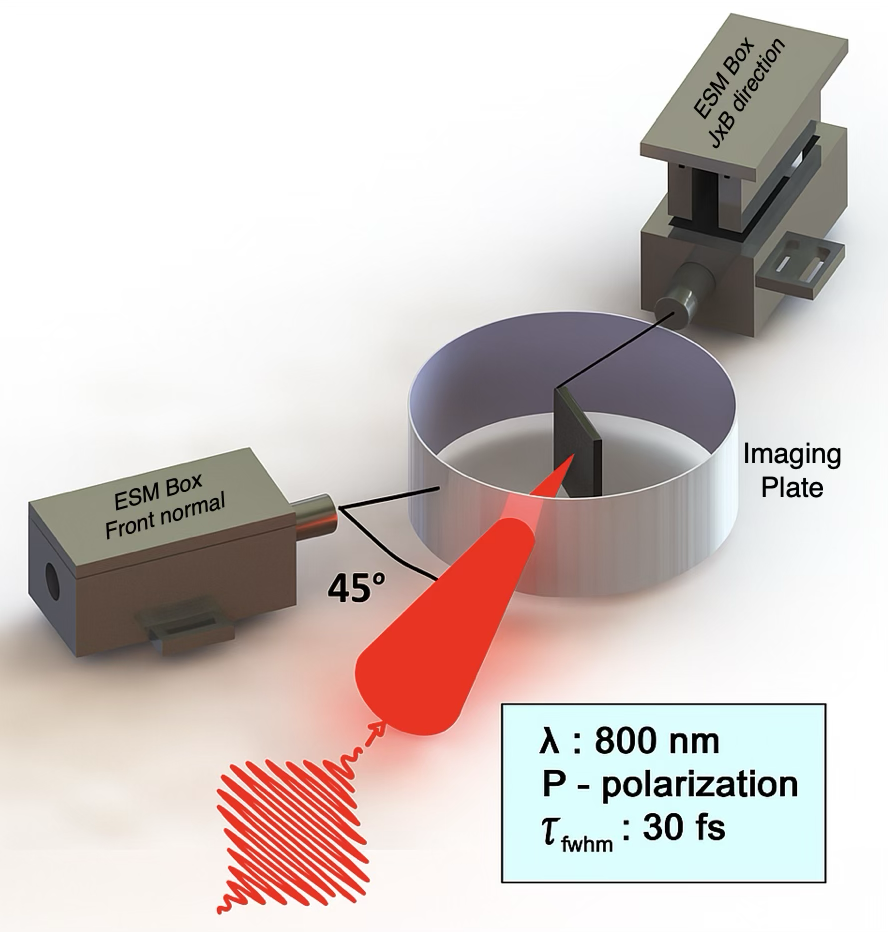}
   \caption{Electron energy spectrum (ESM) and angular distribution measurement schematic}
   \label{fig:Schematic}
\end{figure}
\begin{figure*}[!ht]
    \centering
    \includegraphics[width=01\linewidth]{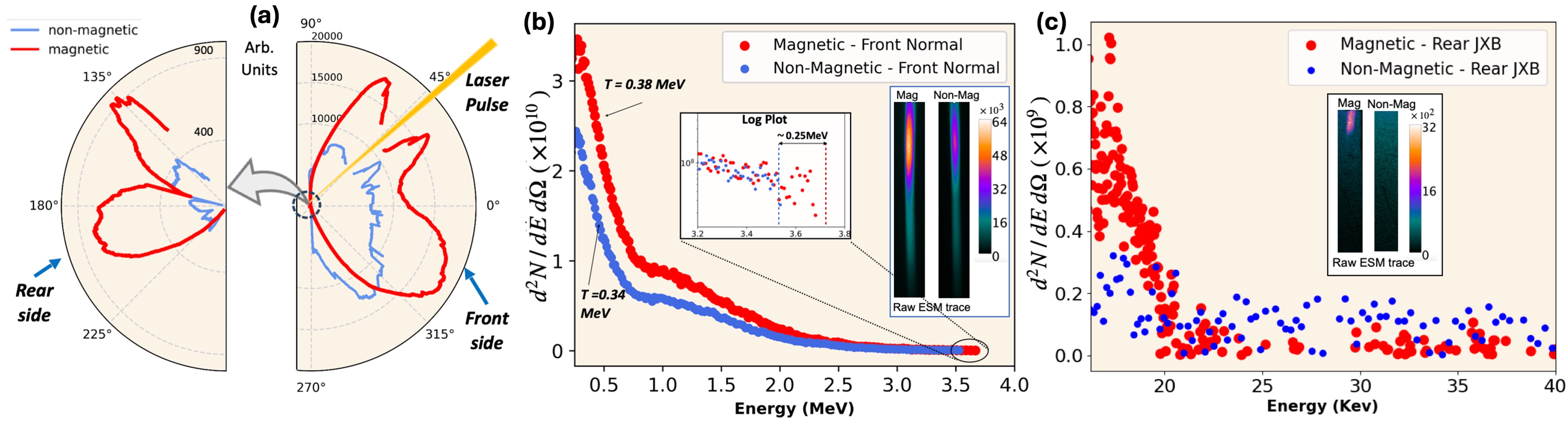}
    \caption{(a) Electron angular distributions for magnetic and non-magnetic targets from experiments. (b) Electron energy spectrum at the front normal side for the magnetic and non-magnetic targets. The inset (log scale) shows an additional 0.25 MeV of energy cut-off for the magnetic target case. ESM traces are also shown (c) Rear side electron energy spectrum with ESM traces of electrons. The non-magnetic target shows no clear signal (mostly noise and background) on the rear side; however, the magnetic target shows a clear signal.}
    \label{fig:results}
\end{figure*}

\section{Experimental Feasibility}\label{sec:experiments}
An important question that naturally arises is the experimental feasibility of the proposed mechanism. The laser parameters employed in the simulations are well within the capabilities of modern high-intensity laser facilities and are routinely accessible in several research laboratories worldwide. At first sight, however, the requirement of external magnetic fields in the kilotesla regime may appear to pose a significant experimental challenge. Recent experimental results from the Tata Institute of Fundamental Research provide a promising avenue in this regard. In these experiments, a $150~\mathrm{TW}$ laser system was used to irradiate inexpensive neodymium permanent-magnet targets ($5~\mathrm{mm}$ thick) possessing an initial axial magnetic field of approximately $0.1~\mathrm{T}$. The targets were exposed to $30~\mathrm{fs}$, $800~\mathrm{nm}$ laser pulses at an intensity of about $10^{19}~\mathrm{Wcm^{-2}}$. Remarkably, the interaction was found to trigger a dynamo-like amplification of the magnetic field by nearly four orders of magnitude, resulting in self-generated magnetic fields approaching $2~\mathrm{kT}$ \cite{choudhary202510}. This observation demonstrates that kilotesla-scale magnetic fields can emerge naturally during laser–target interactions, even when starting from relatively modest seed fields. Indeed, this experimental finding served as one of the primary motivations for choosing external magnetic fields of the order of a few kilotesla in our simulations. It suggests that the parameter regime required for the resonant acceleration mechanism proposed here may already be accessible using relatively simple target configurations and existing table-top high-power laser systems.

Therefore, the post-interaction plasma dynamics corresponding to a $2~\mathrm{kT}$ magnetized target can be effectively realized in this experimental configuration, providing an opportunity to directly test the predictions of our simulations. As illustrated in Fig.~\ref{fig:Schematic}, the energies, fluxes, and angular distributions of the fast electrons are measured at both the front and rear surfaces of the $5~\mathrm{mm}$-thick target. 
Further details of the experimental setup and measurement techniques are provided in the Supplementary Material.

The experimental observations clearly demonstrate that the presence of a magnetic field leads to a substantial enhancement in the energy of the emitted electrons, in qualitative agreement with the trends predicted by the simulations. Owing to the time-integrated nature of the present diagnostics, however, the measurements are unable to resolve the quasi-monoenergetic spectral features predicted by the numerical studies. Future experiments should therefore focus on the development of time-resolved and high-resolution electron diagnostics capable of capturing the transient formation of these spectral peaks. In parallel, efforts are required to further strengthen the monoenergetic feature itself. This may be achieved by optimizing the excitation of electron Bernstein waves and enhancing their subsequent Landau damping \cite{bernstein1958waves,Laqua2007}, thereby increasing the efficiency of resonant energy transfer to the electron population. Such advances could pave the way toward the realization of highly controllable, high-flux, quasi-monoenergetic electron sources based on magnetized overdense plasma interactions.


Figures \ref{fig:results}(a) depict the angular distribution of electron fluxes observed in the experimental data, respectively, from both the front and rear surfaces of the target. Only fast electrons with energies exceeding $100$ keV were detected. The figures clearly demonstrate that the magnetized target exhibits a much higher electron flux than the non-magnetized target. However, the angular distribution of electrons remains relatively broad at both the front and back surfaces of the target.
The ESM results, shown in Figure \ref{fig:results}(b), also exhibit similar trends. The effective temperature of hot electrons is higher in a magnetized target than in a non-magnetized one. The cutoff energy on the tails of the spectrum is also higher ($\sim$3.7 MeV compared to 3.5 MeV) for the magnetized target. Figure \ref{fig:results}(c) displays the energy spectrum for both targets at the rear. No distinct signal (noise and background only) is observed for the non-magnetic target, while the magnetic target exhibits a clear and enhanced signal above the noise level. This represents the unmistakable signature of energetic electrons effectively traversing the 5mm-thick Neodymium magnet, enduring energy loss due to collisions and other processes in the presence of an external static magnetic field.

\begin{table*}[htbp]
\centering
\caption{Comparison of electron acceleration performance.}
\label{tab:comparison}

\resizebox{\textwidth}{!}{
\begin{tabular}{|c|l|c|c|c|c|c|c|c|c|}
\hline
Year &
Reference &
Laser Power &
$n_e$ (cm$^{-3}$) &
$n_e/n_c$ &
Accel. Length &
Charge (pC) &
Energy (GeV) &
$E/L$ (MeV/$\mu$m) &
Energy Spread \\
\hline

2013 &
Wang et al.\cite{wang2013quasi} &
1 PW &
$5\times10^{17}$ &
$2.9\times10^{-4}$ &
$\sim 2$ cm &
$\sim 63$ &
2.0 &
0.10 &
$\sim 10\%$ \\
\hline

2014 &
BELLA \cite{leemans2014multi} &
0.3 PW &
$5\times10^{17}$ &
$2.9\times10^{-4}$ &
$\sim 4$ cm &
$\sim 10$ &
4.2 &
0.105 &
$\sim 6\%$ \\
\hline

2021 &
Salehi et al. \cite{salehi2021laser}&
0.5 TW &
$(1-1.7)\times10^{20}$ &
0.08--0.13 &
0.013 cm &
2.5 &
0.015 &
0.115 &
$\sim 20\%$ \\
\hline

2025 &
Meter-scale PWFA \cite{zhang2025plasma} &
10 GeV beam &
$(1.3-3.2)\times10^{17}$ &
-- &
125 cm &
2.3 &
26 &
0.013 &
$\sim 0.7\%$ \\
\hline

\textbf{2026} &
\textbf{This work (Magnetized Solid Plasma)} &
150 TW &
$1.747\times10^{23}$ &
100 &
538 nm &
1.54 &
0.002 &
\textbf{3.6} &
$\sim 6.7\%$ \\
\hline

\end{tabular}
}
\end{table*}

These experimental observations are in good agreement with our simulation results. At the front surface of the target, a large population of electrons is extracted during the laser–plasma interaction, generating a strong, directed sheath field, as shown in Fig.~\ref{fig:sheath_fluctuation}(a). This sheath field, in conjunction with the excited electron Bernstein mode, leads to the formation of quasi-monoenergetic electron populations at higher energies. As illustrated in Fig.~\ref{fig:particle_trajectory}(a), these energetic electrons propagate through the target. Although the target thickness (5 mm) causes substantial energy loss, a fraction of the electrons successfully traverse the material and produce a bright spot in ESM at the rear surface along the $\vec{J}\times \vec{B}$ direction. Therefore, these results confirm that a even in weak magnetized Nd target, delayed magnetization effect arises through the generation of quasi-monoenergetic electron population. 

\section{Conclusions and Broader implications}
Table \ref{tab:comparison} places the present results in the context of representative laser-plasma acceleration experiments reported in the literature. While conventional laser wakefield acceleration schemes have achieved impressive electron energies extending from the MeV to multi-GeV regime, they rely on laser propagation through highly underdense plasmas, typically with densities several orders of magnitude below the critical density. This requirement fundamentally limits the plasma density available for acceleration, thereby constraining the achievable beam charge and current.

The present work demonstrates a fundamentally different acceleration paradigm operating in a magnetized overdense plasma with $n_e/n_c=100$. Through particle-in-cell simulations, we have shown that the interaction of an ultraintense laser pulse with a magnetized overdense target leads to the emergence of quasi-monoenergetic electron populations superimposed on a broad thermal background. The mechanism is shown to originate from the excitation of electron Bernstein waves at the plasma boundary and their subsequent Landau damping, which selectively transfers energy to electrons satisfying a relativistic cyclotron-resonance condition. The resulting electron beams are directional, tunable through the applied magnetic field, and generated without requiring laser propagation into the plasma bulk.

Although the electron energies obtained in the present proof-of-principle study remain in the MeV range, the acceleration gradient reaches approximately $\sim3.6~\mathrm{MeV/\mu m}$, exceeding by more than an order of magnitude the values reported in the representative conventional schemes listed in Table \ref{tab:comparison}. This remarkable gradient is achieved over a characteristic acceleration length of only $\sim 538~\mathrm{nm}$ and highlights the potential of overdense plasma environments as compact high-field accelerators. Furthermore, recent experimental observations of kilotesla magnetic-field generation from simple permanent-magnet targets suggest that the physical regime explored here is experimentally accessible with existing high-intensity laser facilities.

It is important to emphasize that the present results represent only an initial exploration of this new regime. No systematic optimization of the plasma density profile, sheath-field structure, target geometry, magnetic-field configuration, or laser parameters has yet been carried out. Since the proposed mechanism relies on the excitation and damping of electron Bernstein waves, substantial improvements may be expected through engineering of the plasma-vacuum interface. In particular, optimization of the sheath thickness and density scale length could enhance the coupling of laser energy into the Bernstein-wave mode, while appropriate tailoring of the magnetic field and laser incidence conditions may increase the wave amplitude and improve the efficiency of its Landau damping. Such optimization is expected to strengthen the monoenergetic feature, increase the accelerated charge, and extend the electron energies to substantially higher values.

From an experimental perspective, future efforts should focus on the development of time-resolved diagnostics capable of directly resolving the quasi-monoenergetic peaks predicted by the simulations, as well as on direct measurements of the electron Bernstein waves responsible for the acceleration process. A systematic exploration of parameter space may reveal operating points with significantly enhanced beam quality and conversion efficiency.

The results presented here, therefore, establish electron Bernstein wave-mediated acceleration in magnetized overdense plasmas as a promising new direction for laser-driven particle acceleration. More broadly, they introduce a new framework in which warm-plasma electrostatic modes, rather than propagating electromagnetic wakefields, act as the intermediary for transferring laser energy to energetic particle beams. This opens exciting opportunities for the development of compact, high-gradient, high-flux electron sources and for the exploration of previously inaccessible regimes of laser-plasma interaction.

Beyond the immediate context of electron acceleration, the present work highlights the possibility of exploiting warm-plasma electrostatic modes as efficient mediators of laser energy coupling in dense plasma environments. This perspective differs fundamentally from conventional acceleration concepts based on propagating electromagnetic wakefields and suggests a broader framework in which collective plasma modes can be engineered to achieve controlled energy deposition and particle acceleration. The results therefore open new directions not only for compact accelerator development but also for the wider study of laser-driven high-energy-density plasmas.

\textbf{Data Availability}\\
The datasets generated and/or analyzed during the current study are available from the corresponding author upon reasonable request. Custom test-particle analysis codes developed during this study are available from the corresponding author upon reasonable request.\\

\begin{acknowledgments}
AD acknowledges support from  the ANRF core grants  CRG/2022/002782 as well as  J C Bose Fellowship grant ANRF/JBG/2025/000237/PS. GRK thanks the Department of Atomic Energy (DAE) for long term support of this research, at present from the grant “Physics and Astronomy (Project Identification No. RTI4002) Department of Atomic Energy,  Tata Institute of Fundamental Research" and acknowledges partial support from the J C Bose Fellowship grant JBR/2020/00039  of the Anusandhan  National Research Foundation (ANRF), both agencies of the Government of India. AD and
TD would like to acknowledge the OSIRIS Consortium,
consisting of UCLA and IST (Lisbon, Portugal) for
providing access to the OSIRIS framework which is
the work supported by NSF ACI-1339893. TD would like to thank IIT Delhi HPC facility for computational resources and the Council for Scientific and Industrial Research (Grant No-09/086/(1489)/2021-EMR-I) for funding the research. ADL acknowledges support from the Infosys-TIFR Leading Edge Research Grant (Cycle 2).

\end{acknowledgments}

\bibliography{ref}

\end{document}



\maketitle

\section{Experimental details}

The experiment was conducted with the Tata Institute of Fundamental Research 150 TW laser system. 5mm-thick neodymium targets (magnetized and unmagnetized) were irradiated with 800 nm, 30-femtosecond p-polarised laser pulses, focused to a $6\mu m$ spot (FWHM) by an $f/3$ off-axis parabolic mirror at $45^{\circ}$ incidence angle. Peak intensity of $8\times$ 10$^{19}$ W/cm$^{2}$ was achieved with $10^{-7}$ intensity contrast at 20 picoseconds. The magnetic field strength in the magnetized target was 0.1 tesla, directed normal to the surface.

To diagnose the distribution of electron fluxes, image plates (IP) were placed in a cylindrical geometry surrounding the target at the center (see Fig~\ref{fig:Exp_Setup}). 10 layers of $11\mu m$-thick aluminum foil are used to cover image plates, blocking low-energy electrons (up to $100 keV$) and other visible emissions. Image plates work on the principle of photo-stimulated luminescence and record the electron flux information, which can be read using image scanners.

We used electron spectrometers (ESM) (Fig~\ref{fig:Exp_Setup}) placed in two different directions (front normal and along the laser propagation at the target rear, which is the $\vec{J}\times \vec{B}$ direction) to measure the electron energies and temperatures. Each electron spectrum was obtained by averaging 15 laser shots. The experiment was performed in a vacuum chamber at a pressure of $10^{-5}$ Torr.

\renewcommand{\thefigure}{S1}
\begin{figure*}[!ht]
   \centering
    \includegraphics[width=0.8\linewidth]{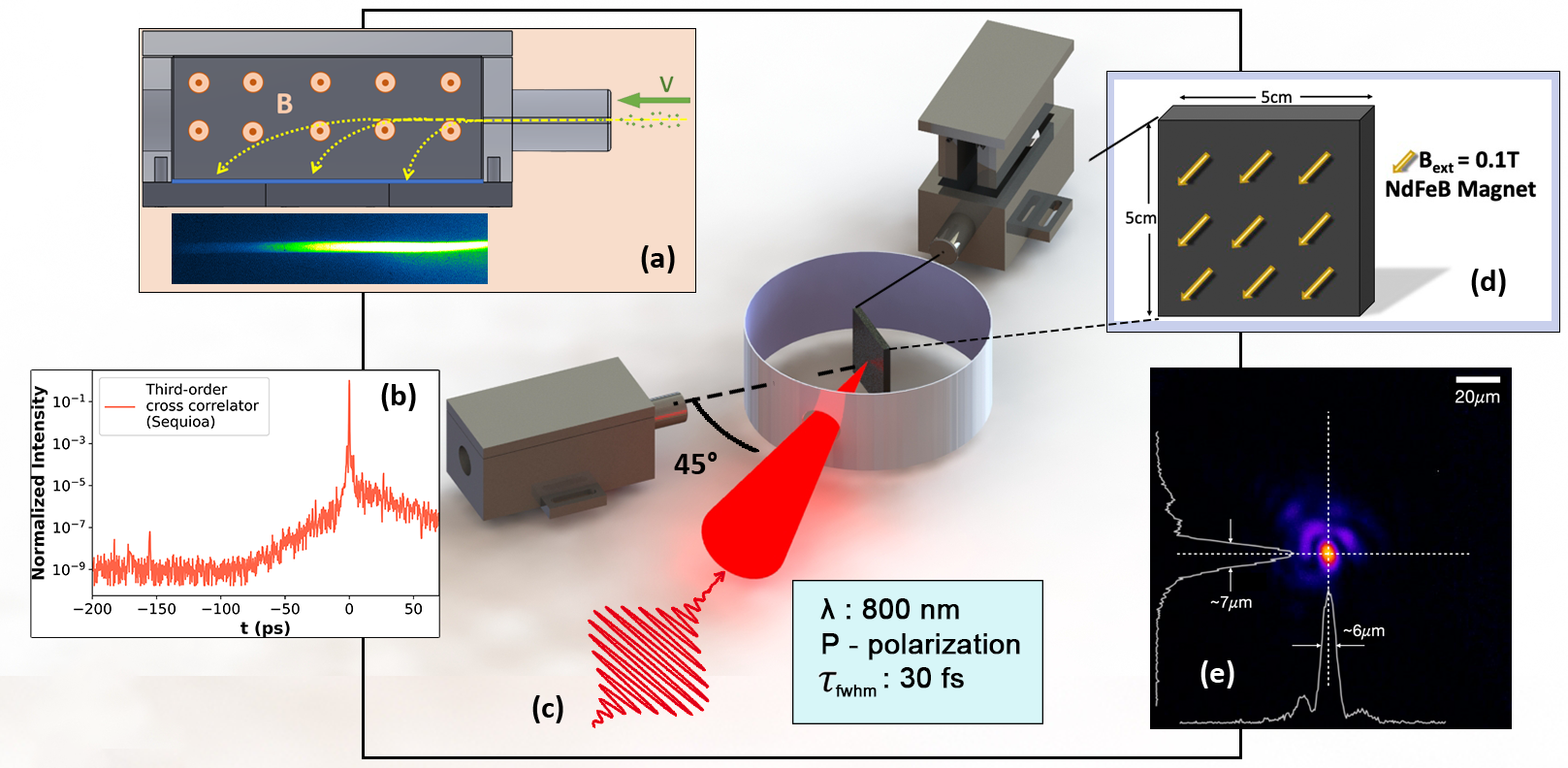}
   \caption{(a) Trajectory of electrons inside ESM(electron spectrometer) magnetic field. The electron trace on the image plate is also shown. (b) contrast measurement of the laser pulse (pulse width - 30fs). (c) Illustration of the ESM and angular distribution measurement setup. (d) Neodymium magnetic target with a 0.1 Tesla magnetic field directed normal to the surface. (e) The laser pulse focal spot (6 $\mu m$ FWHM ) measured during the experiment.}
   \label{fig:Exp_Setup}
\end{figure*}

\bibliography{ref.bib}